\documentclass[letterpaper,10pt]{article} 

\usepackage{osameet3} 

\usepackage{amsmath,amssymb,comment,cite,wrapfig}
\usepackage{algorithm}
\usepackage{algpseudocode}
\usepackage[center,font=footnotesize,labelfont=bf,skip=0.5pt]{caption}
\usepackage[colorlinks=true,bookmarks=false,citecolor=blue,urlcolor=blue]{hyperref}

\usepackage[]{inputenc}
\usepackage{tikz}
\usetikzlibrary{shapes,arrows}

\begin{document}

\title{HeCSON: Heuristic for Configuration Selection in Optical Network Planning}

\author{Sai Kireet Patri\textsuperscript{(1,2)}, Achim Autenrieth\textsuperscript{(1)}, Danish Rafique\textsuperscript{(1)}, J\"org-Peter Elbers\textsuperscript{(1)} and Carmen Mas Machuca\textsuperscript{(2)}}
\address{(1) ADVA,Fraunhoferstr. 9A, Munich, Germany (2) Chair of Communication Networks, Technical University of Munich, Arcisstr. 21, Munich, Germany}
\email{SPatri@advaoptical.com}


\begin{abstract}
We present a transceiver configuration selection heuristic combining Enhanced Gaussian Noise (EGN) models, which shows a 40\% increase in throughput and 87\% decrease in execution time, compared to only approximate EGN and Full-Form EGN respectively.
\end{abstract}
\vspace{-0.3 em}
\vspace{-0.25em}
\section{Introduction}
\vspace{-0.5em}
With advances in transponders supporting software tunable channel configurations \cite{adva19, cisco2019}, network planners possess a new degree of freedom to select the best \textbf{configuration}, i.e. combination of data rate, modulation format and forward error correction (FEC) for each light-path. In-house network planning tools and open-source software like \cite{auge2019}, require planners to manually enter this information. However, when multiple combinations of modulation formats (QPSK, 8QAM, 8/16QAM, 32QAM, 64QAM), data rate (100-600 Gbps) and FEC (0\%,15\%,27\%) are possible, manual configuration assignment becomes tedious. Since non-linear noise in a fiber depends on path parameters like neighbouring channels, span lengths, etc., choosing configurations for each light-path  affects the Quality of Transmission (QoT) in the network. Planners also need knowledge of the best possible data rate for a source-destination pair, while undertaking network feasibility studies, in order to provide guarantee of service. Hence, there is a need for an algorithm which quickly provides a list of viable configurations for each demand, considering both linear and non-linear interference. We define a demand as a single bi-directional light-path between a source-destination pair, which could be assigned any configuration.\\ 
For a QoT metric to measure the performance of each channel, typically Optical Signal to Noise Ratio (OSNR) or in-house system engineering rules are used. In our work, we use OSNR and define two types, namely \textbf{linear OSNR}, i.e. $\frac{P_{ch}}{P_{ASE}}$, where $P_{ch}$ is the channel launch power, $P_{ASE}$ is the Amplified Spontaneous Emission (ASE) power (12.5 $GHz$ noise bandwidth); and \textbf{total OSNR}, i.e., $\frac{P_{ch}}{P_{ASE}+P_{NLI}}$, where $P_{NLI}$ is the non-linear interference (NLI) noise power. To model $P_{NLI}$, the Full-Form Enhanced Gaussian Noise Model (FF-EGN) \cite{carena14}, has emerged as a more accurate model as compared to the Incoherent Gaussian Noise Model among other variants \cite{poggiolini2012detailed}. However, FF-EGN is calculation intensive and does not scale well in terms of time and resources for large networks with many A-Z demands. To cater for this, \cite{ranjbar2019} presents an approximate closed-form EGN formula (ACF-EGN) , which shows good results in end-to-end links, but has not been verified on heterogeneous network topologies.\\
FF-EGN has recently been used in \cite{rabbani2019} to maximize a network's achievable rate by optimizing the launch power of 200 Gbps based configurations. However, the assumption of homogeneous span lengths and lack of configurations with higher data rates does not apply to heterogeneous networks.\\
We propose a ``Heuristic for Configuration Selection in Optical Networks" (\textbf{HeCSON}), which encompasses both the quick and approximate ACF-EGN and the slow but accurate FF-EGN to find the best available configuration for a given demand, and apply it to networks with heterogeneous span lengths, variable gain amplifiers and multiple data rates.
\vspace{-0.75em}
\section{HeCSON and Path Calculation Tests}
\vspace{-0.3em}
\subsection{Network Information and Configurations}
\vspace{-0.3em}
HeCSON is run on the Germany50 network (50 nodes, 88 links, 662 demands), Nobel-EU network (28 nodes, 41 links, 378 demands) and Norway network (27 nodes, 50 links, 351 demands) \cite{sndlib19}. Since there is no open source physical network information available for these networks, we have modelled the length of spans and type of amplifiers from distributions of real networks. For reproducible results, these physical network information files will be made available upon publication.\\
In order to achieve ``new" configurations like 600G 32/64QAM with 15\% FEC \cite{acacia2018}, we start by creating various combinations of data rates between $100$ to $600$ Gbps, in steps of 50 Gbps, bits per symbol from $2$ to $6$, and FEC of $0\%$, $15\%$ or $27\%$. From these, we select the configurations with symbol rate between $32$ and $72$ GBaud, which leaves us with \textbf{60 possible configurations}. The worst case transceiver performance, given by the \textbf{minimum required OSNR}, is available for some configurations in \cite{cisco2019}. We extrapolate these and assign a minimum required OSNR to each configuration, which is then used as a selection and validation metric in HeCSON. 

\vspace{-0.5em}
\subsection{HeCSON Workflow}\label{Sec:heuristic}
\vspace{-0.75em}
\tikzstyle{decision} = [diamond, draw, fill=blue!40, 
    text width=1.0em, text badly centered, node distance=2cm, inner sep=0pt]
\tikzstyle{decision_green} = [diamond, draw, fill=green!30, 
    text width=2.75em, text badly centered, node distance=2cm, inner sep=0pt]
\tikzstyle{decision_orange} = [diamond, draw, fill=orange!60, 
    text width=2.5em, text badly centered, node distance=2cm, inner sep=0pt]
\tikzstyle{block} = [rectangle, draw, fill=blue!15, 
    text width=4em, text centered, rounded corners, minimum height=1em]
\tikzstyle{block_presel} = [rectangle, draw, fill=blue!15, 
    text width=4em, text centered, rounded corners, minimum height=1em]
\tikzstyle{block_green} = [rectangle, draw, fill=green!30, 
    text width=6em, text centered, rounded corners, minimum height=1em]
\tikzstyle{block_green_acf} = [rectangle, draw, fill=green!30, 
    text width=3.5em, text centered, rounded corners, minimum height=2em]
\tikzstyle{block_purple} = [rectangle, draw, fill=purple!30, 
    text width=3em, text centered, rounded corners, minimum height=1em]
\tikzstyle{block_orange} = [rectangle, draw, fill=orange!60, 
    text width=3em, text centered, rounded corners, minimum height=1em]
\tikzstyle{line} = [draw, -latex']
\tikzstyle{cloud} = [draw, ellipse,fill=red!20, text badly centered,minimum height=0.25em]
\tikzstyle{every node}=[font=\scriptsize]

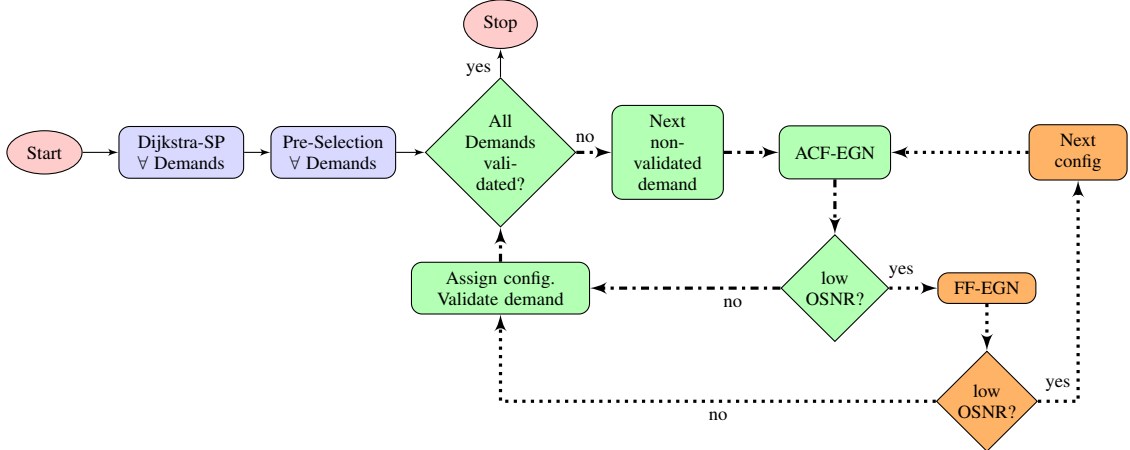
\begin{figure}[htbp!]
    \centering
\vspace{-0.75em}
\begin{tikzpicture}[node distance = 0.5cm, auto]
    
    \node [cloud] (demands) {Start};
    \node [block,right of =demands,node distance=1.8cm] (dijk) {Dijkstra-SP $\forall$ Demands};
    \node [block_presel, right of=dijk,node distance=2cm] (pre-sel) {Pre-Selection $\forall$ Demands};
    \node [decision_green, right of=pre-sel,node distance=2.2cm] (decideDem) {All\\Demands validated?};
    \node [block_green_acf,right of =decideDem,node distance=2.2cm] (next) {Next\\non-validated demand};
    \node [block_green_acf, right of=next,node distance=2.2cm] (configsel) {ACF-EGN};
    \node [cloud, above of=decideDem,node distance=1.7cm] (stop) {Stop};
    \node [decision_green, below of=configsel,node distance=1.8cm] (decideApprox) {low OSNR?};
    \node [block_orange, right of=decideApprox, node distance=2cm] (validation) {FF-EGN};
    \node [block_green, below of=decideDem, node distance=1.8cm] (assign) {Assign config. Validate demand};
    \node [decision_orange,below of=validation, node distance=1.5cm](decideFull){low OSNR?};
    \node [block_orange, right of=configsel, node distance=3.2cm] (newConfig) {Next config};
    
    \path [line] (dijk) -- (pre-sel);
    \path [line] (pre-sel) -- (decideDem);
    \path [line] (decideDem) -- node[near start] {yes} (stop);
    \path [line,very thick,dash dot] (decideDem) -- node[near start] {no} (next);
    \path [line,very thick,dash dot] (next) -- (configsel);
    \path [line,very thick,dash dot] (configsel) -- (decideApprox);
    \path [line,very thick,dash dot] (decideApprox) -- node[near start] {no} (assign);
    \path [line,very thick, dotted] (decideApprox) -- node[near start] {yes} (validation);
    \path [line,very thick, dotted] (validation) -- (decideFull);
    \path [line,very thick, dotted] (decideFull) -| node[near start] {yes} (newConfig);
    \path [line,very thick, dotted] (decideFull) -| node[near start] {no} (assign);
    \path [line,very thick,dash dot] (assign) -- (decideDem);
    \path [line,very thick,dotted] (newConfig) -- (configsel);
    \path [line] (demands) -- (dijk);
    
\end{tikzpicture}
\caption{HeCSON's Flowchart}
\label{fig:flowchart}
\end{figure}
\vspace{-1.25em}

Using Fig. \ref{fig:flowchart}, we explain the heuristic. Starting with the list of demands and a list of all possible configurations, we calculate Dijkstra's shortest path (Dijkstra-SP) for all the demands and then perform \textbf{Pre-Selection} by calculating the linear OSNR for all demands based on the demand length. The configurations for which the linear OSNR drops below the minimum required OSNR are removed from the available configuration list of a particular demand. The configuration list is then ordered by first maximizing data rate and then minimizing channel bandwidth amongst the ordered maximized data rates. The first available configuration for each demand is then taken from the list and is placed using a first fit channel allocation.\\ Once all demands are allocated an initial configuration, we iterate over each demand to check if it is validated and begin \textbf{Configuration Selection}. After selecting the next non validated demand, ACF-EGN block calculates the total ACF-EGN OSNR and checks if it is higher than the minimum required OSNR of the assigned configuration (low OSNR with dash-dot lines). In case it is, the configuration is assigned to the demand, otherwise, we execute \textbf{Configuration Validation} (blocks connected with dotted lines in Fig. \ref{fig:flowchart}). For the current configuration, the total FF-EGN OSNR is calculated (FF-EGN block). If that too falls below the minimum required OSNR (low OSNR connected with dotted lines), the configuration is removed from the channel allocation matrix and the next configuration available for the demand is placed (Next config). Otherwise, in case of validation, we assign the configuration to the demand and mark it as ``validated" for HeCSON. The demands for which no configuration is selected due to not satisfying the minimum OSNR requirements are blocked.
\vspace{-0.7em}
\subsection{Path calculation tests}
\vspace{-0.5em}
We compare HeCSON with a configuration selection based on a pure ACF-EGN model and a pure FF-EGN model. In both these ``pure" cases, every time a placed configuration's total OSNR falls below the minimum required total OSNR, the next configuration is placed and the total OSNR is re-calculated. The two major parameters we evaluate in our study are the \textbf{total network throughput}, defined as the sum of all the placed demand data rates and the \textbf{total execution time}, defined as the time taken for every case to run using a Java program on an Intel Core i7 processor with 16 GB RAM. The other results of interest are \textbf{number of blocked demands}, which are the demands blocked either due to lack of spectrum slots or suitable configuration; and \textbf{spectral efficiency}, defined as the ratio of overall network throughput ($Tbps$) to the total occupied spectrum on all links in the network ($THz$). It should be noted that we consider throughput results only for those demands which are placed by all three cases. We look at the throughput to show that the use of a multi-configuration, flexible-grid transceivers in the network provide higher data rates for demands. Execution time is important for future online planning applications, so that the algorithm can do real-time OSNR recalculation for all neighbouring channels whenever a new demand is added.
\begin{figure}[htbp!]
\vspace{-1.0em}
    \centering
    \includegraphics[width=\textwidth]{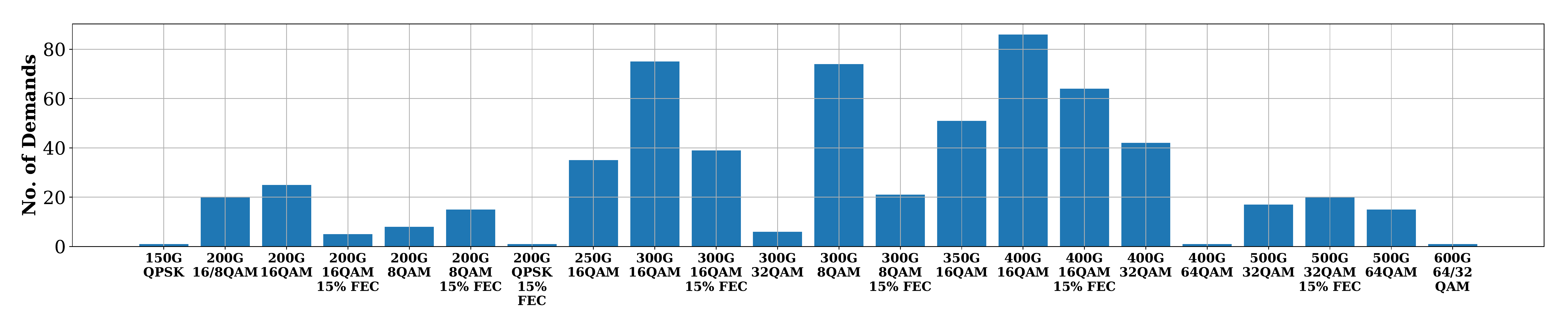}
    \caption{Distribution of demands for each of the 22 selected configurations in the Germany50 network by HeCSON}
    \vspace{-2em}
    \label{fig:plotConfigHist}
\end{figure}
\vspace{-0.5em}
\section{Results and Discussion}
\begin{figure}[htbp!]
\vspace{-1.75em}
    \centering
    \includegraphics[width=\textwidth]{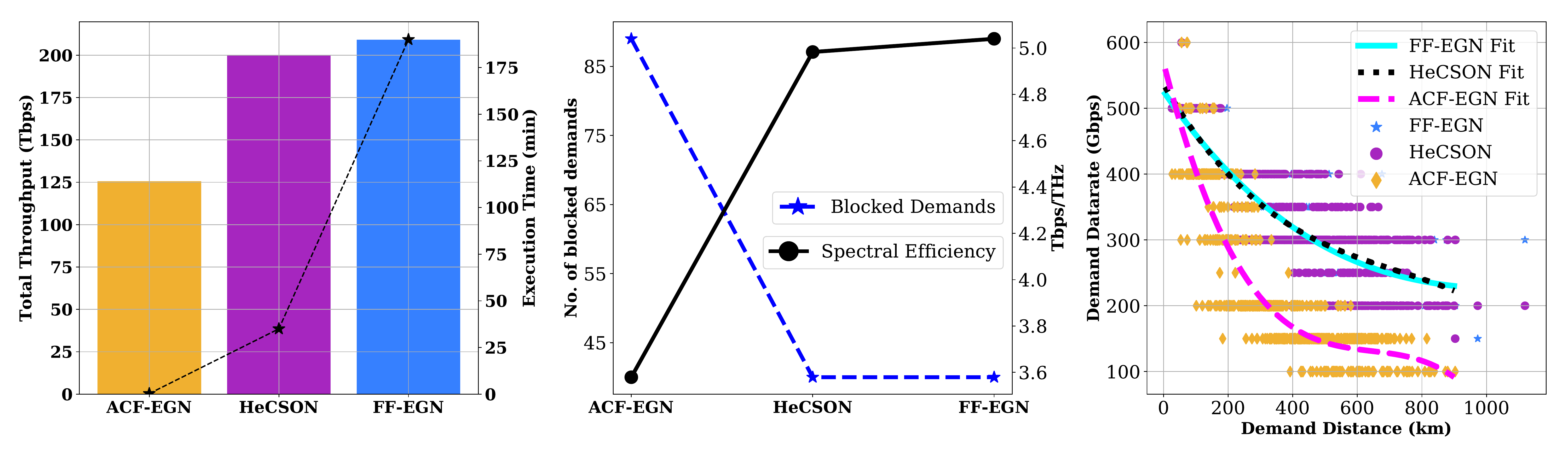}
    \vspace{-0.5em}
    \caption{For the Germany50 network, \textbf{(a)} Network Throughput and Execution Time. \textbf{(b)} Number of blocked demands and spectral efficiency (\textbf{c}) Polynomial fit of data rate vs demand distance for all three cases}
    \vspace{-1.5em}
    \label{fig:demandDatarateSubPlots}
\end{figure}

\vspace{-0.5em}
Upon running HeCSON for Germany50 network, we see that only 22 of the 60 possible configurations have been used. This is shown in Fig. \ref{fig:plotConfigHist}, where majority of the demands are placed between 200 and 400 Gbps configurations. We see a substantial number of demands selecting 300 and 400 Gbps 15\% FEC based configuration, which provide higher spectral efficiency compared to 27\% FEC based configurations. The application of our pre-filtering step, as explained in Section \ref{Sec:heuristic}, helps in reducing the execution time vastly and removing configurations for which even the linear OSNR would fall below the minimum required OSNR.
From Fig. \ref{fig:demandDatarateSubPlots} (a), we see that HeCSON provides a total network throughput of 199.85 Tbps, which is comparable to 209.20 Tbps suggested by the FF-EGN case.\\

HeCSON completes execution in 35 minutes (star plot in Fig. \ref{fig:demandDatarateSubPlots} (a)), compared to 190 minutes for FF-EGN case. Reduction in time is helpful for planners to test out different scenarios and configurations quickly. Comparing with ACF-EGN case, the overall network throughput is increased by more than 75 Tbps.\\For further generalization of our analysis, we also look at the spectral efficiency (Fig. \ref{fig:demandDatarateSubPlots} (b)) for each of the cases. We see that HeCSON has a spectral efficiency of 5.02 $Tbps$/$THz$, comparable with FF-EGN case's spectral efficiency of 5.04 $Tbps$/$THz$ and having a 60\% increase as compared to ACF-EGN.
\begin{wrapfigure}[12]{r}{0.4\textwidth}
\vspace{-1.75em}
    \centering
    \includegraphics[width=0.4\textwidth]{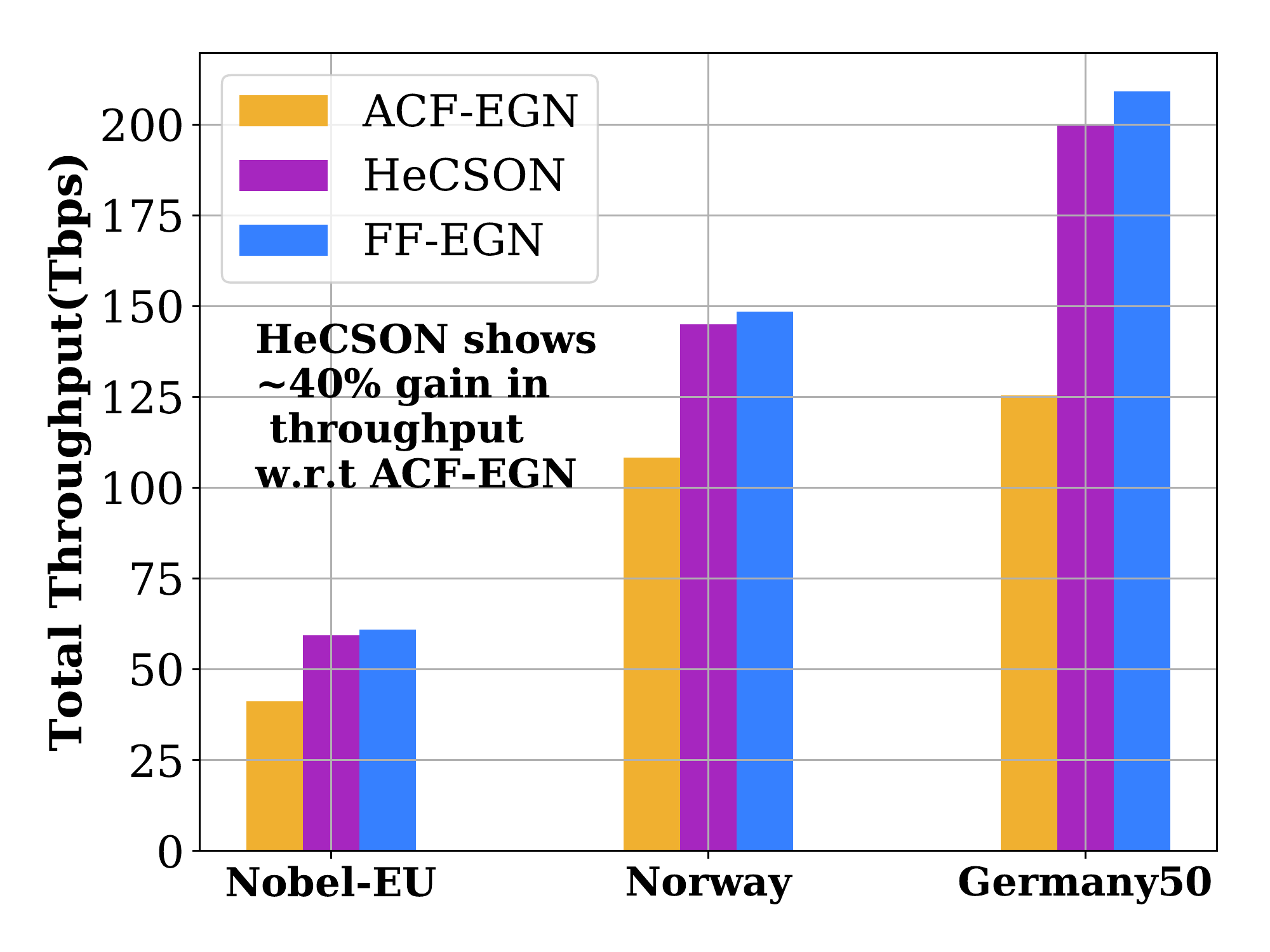}
    \vspace{-1.75em}
    \caption{Total throughput for three networks}
    \vspace{-1.75em}
    \label{fig:plotNW}
\end{wrapfigure}
For all three cases, at least 40 demands are blocked in the Pre-Selection stage, due to lack of adequate spectrum slots on some links. Additional 49 demands are blocked in ACF-EGN case due to not satisfying the minimum required OSNR. This is because ACF-EGN, like other approximations mentioned in \cite{carena14}, overestimates $P_{NLI}$. Looking at the chosen data rate for each demand, we see a clear decrease in data rate, as the distance increases. When fitted using a third-order polynomial, we see that HeCSON's fitted curve closely follows FF-EGN's curve. Running the path calculation tests on different sized networks, we see from Fig. \ref{fig:plotNW} that the throughput follows similar trends in each of them. For the three networks, namely, Nobel-EU, Norway and Germany50, we consider maximum possible bi-directional demands between non-adjacent source-destination pairs, i.e., 378, 351 and 662 respectively. Compared to ACF-EGN, HeCSON provides throughput gain of approximately 40\%. We can therefore conclude that HeCSON is network agnostic and can be used in different optical networks for transceiver configuration assignment.\\To summarize, we propose HeCSON as a new heuristic for configuration selection in heterogeneous optical networks. About 40\% gain in the total network throughput as compared to the conservative pure ACF-EGN case and an 87\% decrease in execution time as compared to the accurate pure FF-EGN case is obtained.\\
\vspace{-1.25em}
 {\scriptsize
\\
This work is partially funded by Germany's Federal Ministry of Education and Research under project OptiCON (grant IDs \textbf{\#16KIS0989K} and \textbf{\#16KIS0991}).
}

\end{document}